\documentclass[journal,transmag]{IEEEtran}
\usepackage{hyperref,color,breqn,multirow}
\usepackage[top=0.7in, left=0.65in]{geometry}
\usepackage{color}
\usepackage{graphicx}
\hypersetup{hidelinks}
\usepackage{amssymb}
\usepackage[T1]{fontenc}
\usepackage{siunitx}
\usepackage{nicefrac}
\usepackage{mathtools}
\usepackage{eso-pic}

\begin{document}
\title{\fontsize{17pt}{17pt}\selectfont
Multi-Level Monte Carlo sampling with
Parallel-in-Time Integration for Uncertainty Quantification in Electric Machine Simulation
}
\author{\IEEEauthorblockN{Robert Hahn \IEEEauthorrefmark{1}, Sebastian Schöps \IEEEauthorrefmark{1}}
\vspace{5pt}
\IEEEauthorblockA{\IEEEauthorrefmark{1}Technical University of Darmstadt, Darmstadt, Germany, robert.hahn@tu-darmstadt.de}
}
\markboth{MLMC with PinT for UQ in Electric Machine Simulation}%
{MLMC with PinT for UQ in Electric Machine Simulation}

\IEEEtitleabstractindextext{%
\begin{abstract}

While generally considered computationally expensive, Uncertainty Quantification using Monte Carlo sampling remains beneficial for applications with uncertainties of high dimension.
As an extension of the naive Monte Carlo method,
the Multi-Level Monte Carlo method reduces the overall computational effort, but is unable to reduce the time to solution in a sufficiently parallel computing environment.
In this work, we propose a Uncertainty Quantification method combining Multi-Level Monte Carlo sampling and Parallel-in-Time integration for select samples, exploiting remaining parallel computing capacity to accelerate the computation. While effective at reducing the time-to-solution, Parallel-in-Time integration methods greatly increase the total computational effort.
We investigate the tradeoff between time-to-solution and total computational effort of the combined method, starting from theoretical considerations and comparing our findings to two numerical examples.
There, a speedup of $12$ -- $45\%$ compared to Multi-Level Monte Carlo sampling is observed, with an increase of $15$ -- $18\%$ in computational effort.
\end{abstract}

}

%arxiv command
\AddToShipoutPicture*{
    \footnotesize\sffamily\raisebox{0.8cm}{\hspace{1.4cm}\fbox{
        \parbox{\textwidth}{
© 2026 IEEE. Personal use of this material is permitted. Permission from IEEE must be
obtained for all other uses, in any current or future media, including
reprinting/republishing this material for advertising or promotional purposes, creating new
collective works, for resale or redistribution to servers or lists, or reuse of any copyrighted
component of this work in other works.
            }
        }
    }
}

\maketitle
\IEEEdisplaynontitleabstractindextext%
\IEEEpeerreviewmaketitle%

\section{Introduction}
Uncertainty quantification (UQ) has become popular in electrical engineering applications such as electrical machine design. In practice, the number of random variables can be substantial, arising from variations in material properties, geometry, boundary conditions and sources~\cite{Clenet_2013aa,Offermann_2013aa,Romer_2016aa}. A common objective is to propagate these uncertainties forward in order to evaluate failure rates, stochastic moments (e.g.\ mean values) and sensitivities of quantities of interest. Several strategies have been developed for this purpose. Among these, stochastic collocation and stochastic Galerkin methods, often combined with polynomial chaos expansions~\cite{Xiu_2010aa}, are efficient when the number of uncertain input parameters is small. However, as the dimensionality of the uncertainty increases, these methods eventually become infeasible due to the curse of dimensionality. While sparse grid techniques~\cite{Bungartz_2004aa} can alleviate this problem to some extent, they cannot eliminate it completely.
On the other hand, the Monte Carlo (MC) method is not affected by dimensionality, but is considered too expensive in its naive implementation. Therefore, Multi-Level Monte Carlo (MLMC) sampling has been developed~\cite{Giles_2008aa} to reduce the computational cost of MC sampling.
In the MLMC method, the numerical solution required for each sample is performed at the desired accuracy level for only a few samples.
For most samples, the numerical solution is performed using a coarser discretization, reducing the total computational effort. The time-to-solution depends on the
parallel computing resources available, but is bounded below by the time-to-solution for a single high-accuracy sample.
In the context of electric machines, the method has already been shown to perform well when hierarchically combining different spatial resolutions~\cite{Galetzka_2019aa}.

In this paper, we introduce MLMC with hierarchical refinement in the time domain for electrical machines. This is a significant advance in cases where, for example, long transient simulations are required to obtain the steady state. In such cases, parallel-in-time (PinT) integration methods~\cite{Gander_2015aa}, which leverage parallel computations to reduce the time-to-solution, become attractive. This reduction in time-to-solution, however, comes at the cost of increasing computational effort, i.e.\ energy spent on the computation. However, significant speed-ups were reported in the context of electrical machines, e.g.\ \cite{Takahashi_2013aa,Bast_2020aa}.
We combine MLMC with PinT to reduce the time-to-solution of a UQ estimation in an HPC environment and investigate the tradeoff between total energy consumption and time-to-solution.

The paper is structured as follows.
In Sec.~\ref{hahn:sec_eddy-current} we recap the finite element formulation used in the numerical simulation of electric machines.
Next, we motivate the use of Monte-Carlo type Uncertainty Quantification methods and show the extension to Multi-Level Monte Carlo sampling in Sec.~\ref{hahn:sec_uq}.
After introducing the Parareal algorithm for Parallel-in-Time integration
and discussing the tradeoff between time-to-solution and computational effort in Sec.~\ref{hahn:sec_pint},
we present the proposed method, combining Multi-Level Monte Carlo sampling with Parallel-in-time integration, and discuss the the influence on time-to-solution and computational effort from theoretical considerations in Sec.~\ref{hahn:sec_proposed}.
We apply the proposed method to two different induction machine models and compare the observed behavior to our previous assumptions in Sec.~\ref{hahn:ch_numerical_example}
and conclude the paper in Sec.~\ref{hahn:ch_conclusion}.
\section{Methodology}%
\label{hahn:ch_methods}
\subsection{Electric machine simulation}%
\label{hahn:sec_eddy-current}
Although electromagnetic phenomena are in general described by Maxwell's equations, it is in many cases sufficient to consider a simplified set of equations. When high spatial accuracy is not needed, e.g.\ in early design stages of induction machines, simpler models, e.g.\ due to Steinmetz~\cite{Steinmetz_1897aa} are used. However, for high-fidelity predictions, one typically employs the magneto-quasistatic approximation to obtain the eddy current problem, see e.g.\ \cite{Salon_1995aa}. Using the electromagnetic vector potential $\mathbf{A}$ with $\mathbf{B} = \operatorname{curl}{\mathbf{A}}$, we solve the eddy current problem in $\mathbf{A}$-formulation
\begin{equation}
    \operatorname{curl}{\mu^{-1} \operatorname{curl}{\mathbf{A}}}
    +\sigma \partial_t \mathbf{A} = \mathbf{J}_{\text{s}} \text{,}
    \label{hahn:eq_eddy-current}
\end{equation}
wherein $\mu$ is the (possibly nonlinear) permeability and $\sigma$ the electric conductivity of the material, and $\mathbf{J}_{\text{s}}$ is the source current density.
Using the finite element (FE) method, the unknown $\mathbf{A}$ is discretized in the spatial domain $\Omega$ using a set of basis functions $\mathbf{w}_i \in H(\operatorname{curl}, \Omega)$ and the unknown is approximated as
\begin{equation}
  \mathbf{A}(\mathbf{x}, t) = \sum_i a_i(t) \mathbf{w}_i(\mathbf{x}) \text{,}
\end{equation}
where the coefficients $a_i(t)$ are the time-dependent degrees of freedom (DoF)~\cite{Monk_2003aa}.
Using a Galerkin approach, we arrive at the semi-discrete system
\begin{equation}
  \mathbf{M}\, \frac{\mathrm{d}{\mathbf{a}}}{\mathrm{d}{t}} +
  \mathbf{K} \mathbf{a}
  = \mathbf{r} \,\text{,}
\label{hahn:eq_eddy-current-semidiscrete}
\end{equation}
with the entries of the mass- and stiffness matrix given by
\begin{equation}
  \begin{aligned}
  \left(\mathbf{M}\right)_{ij} &= \int_\Omega \sigma \mathbf{w}_i \cdot \mathbf{w}_j \,\mathrm{d}{\Omega} \,\text{,}\\
  \left(\mathbf{K}\right)_{ij} &= \int_\Omega \mu^{-1} \operatorname{curl} \mathbf{w}_i \cdot \operatorname{curl} \mathbf{w}_j \,\mathrm{d}{\Omega} \,\text{,}\\
  \end{aligned}
\end{equation}
the right-hand-side vector $\mathbf{r}$ given by
\begin{equation}
  \left(\mathbf{r}\right)_i = \int_\Omega \mathbf{J}_{\text{s}} \cdot \mathbf{w}_i \,\mathrm{d}{\Omega} \,\text{,}
\end{equation}
and the DoFs $a_i(t)$ arranged in the time-dependent vector
\begin{equation*}
  \left(\mathbf{a}(t)\right)_i = a_i(t) \text{.}
\end{equation*}
Any variations in the geometry of the model or in the material parameters $\mu$ or $\sigma$ then lead to variations in $\mathbf{M}$, $\mathbf{K}$, or $\mathbf{r}$ and thus in the solution $\mathbf{a}(t)$.
To keep the number of DoFs manageable, a typical simulation workflow exploits existing symmetries whenever possible, by restricting the computational domain to only part of the model and enforcing symmetry via suitable boundary conditions.
In electric machine simulations, the domain can often be restricted to a single pole of the machine.
As uncertainties in the geometry or materials cannot be assumed to influence the model symmetrically, this approach is usually not applicable in the context of UQ, and the full model has to be resolved.

We solve~\eqref{hahn:eq_eddy-current-semidiscrete} via time integration using e.g.\ the implicit Euler scheme.
There, we introduce the discrete time steps $t_k$ and
approximate the time derivative as
\begin{equation}
  \frac{\mathrm{d}{\mathbf{a}}}{\mathrm{d}{t}}(t_{k+1}) \approx \frac{\mathbf{a}(t_{k+1}) - \mathbf{a}(t_{k})}{\Delta t_{k+1}}
\end{equation}
with time step size $\Delta t_{k+1} = t_{k+1} - t_k$.
After successively solving
\begin{equation}
  \mathbf{M}\, \frac{\mathbf{a}(t_{k+1}) - \mathbf{a}(t_{k})}{\Delta t_{k+1}}
  + \mathbf{K} \mathbf{a}(t_{k+1})
  = \mathbf{r}(t_{k+1})
\label{hahn:eq_eddy_current_euler}
\end{equation}
for $\mathbf{a}(t_{k+1})$ starting from the initial value $\mathbf{a}(t_0)$,
field solutions and other quantities can be computed in a post-processing step.
\subsection{Uncertainty Quantification}%
\label{hahn:sec_uq}
Let us assume that the model, i.e.\ the induction machine, depends on input parameters $\boldsymbol{\xi}$ and performing the simulation produces
a numerical solution $\mathbf{u}(t; \boldsymbol{\xi})$ from which some output quantity $Q(\boldsymbol{\xi})$, e.g.\ losses or torque, is computed in a post-processing step. The inputs are seen as realizations of some underlying random distribution and thus $Q$ is random as well. Its stochastic moments, e.g.\ expected value, are then determined by evaluating samples. The Monte Carlo (MC) method draws $N$ independent samples $\boldsymbol{\xi}^{(k)}, k \in \{1, \dots, N\}$ from the input parameter distribution and performs a numerical simulation to determine the corresponding quantity of interest (QoI), that is $Q^{(k)} = Q(\boldsymbol{\xi}^{(k)})$, for each sample.
The MC estimation for the expected value of the output quantity $Q$ is then computed via
\begin{equation}
\mathbb{E}_{\text{MC}}\left[ Q(\boldsymbol{\xi}) \right] = \frac{1}{N} \sum_{k=1}^N Q^{(k)}\text{.}
\end{equation}
The quality of this estimate depends on the number of samples $N$.
As the MC estimator is unbiased, i.e. $\mathbb{E}[\mathbb{E}_{\text{MC}}[Q]] = \mathbb{E}[Q]$, we can expect to recover the true value of $\mathbb{E}[Q]$
for $N \to \infty$, with a Root Mean Square Error (RMSE) of $\mathcal{O}(N^{-0.5})$ which is, as mentioned previously, rather slow but independent of the dimension of  $\boldsymbol{\xi}$.
Note that the true expectation $\mathbb{E}[Q]$ depends on the accuracy of the numerical method used to calculate $Q(\boldsymbol{\xi})$.
In addition to a high sample count $N$, a sufficiently accurate (and thus computationally expensive) numerical method therefore has to be employed to obtain a sufficiently accurate estimate $\mathbb{E}_{\text{MC}}[Q]$.

As all sample evaluations are independent of each other, the numerical solution required for each sample can be performed in parallel without communication overhead. The method is thus well-suited for high-performance computing (HPC) environments, which offer a large number of processor cores $n_{\text{c}}$, typically distributed among many computing nodes.
In the limit case of $n_{\text{c}} \to \infty$, all numerical solutions can be performed in parallel. This brings the time-to-solution of the MC estimation down to the time-to-solution of a single numerical solution.
The computational effort of the estimation --- that is the sum of the time spent over all active processor cores, which is correlated to the amount of energy spent on the computation --- is unaffected by parallel sample evaluation.
Due to the poor convergence rate of MC sampling and the resulting high sample count, the computational effort can be very large.

To reduce the computational effort of MC sampling, the Multi-Level Monte Carlo (MLMC) method introduces a hierarchy of discretization levels $\ell \in \{0, ..., L\}$ where $\ell=L$ corresponds to the desired accuracy, and $\ell \in \{0, ..., L-1\}$ correspond to coarser discretizations, which produce less accurate solutions at a lower computational cost.
To that end, one can employ coarser spatial discretization, e.g.\ a coarser mesh in a finite element context, as was shown in~\cite{Galetzka_2019aa}.
In this work, we employ coarser time discretizations, i.e.\ larger timesteps $\Delta t$ in the time intergation scheme~\eqref{hahn:eq_eddy_current_euler}.

In the MLMC method, $N_\ell$ samples are then drawn for each discretization level $\ell$, and two numerical solutions are performed --- using discretization levels $\ell$ and $\ell-1$.
Samples drawn for $\ell=0$ only need one numerical solution, as we set $Q_{-1} \equiv 0$.
The MLMC estimate of the expected value is then given by
\begin{equation}
\mathbb{E}_{\text{MLMC}}[Q(\boldsymbol{\xi})] =
    \sum_{\ell=0}^L \frac{1}{N_\ell}
    \sum_{k=1}^{N_\ell} \left[Q^{(k, \ell)}_\ell - Q^{(k, \ell)}_{\ell-1}\right]
    \text{,}
    \label{hahn:eq_mlmc-scheme}
\end{equation}
as introduced in~\cite{Giles_2008aa}, wherein $Q_{\ell}^{(k, l)}$ denotes the value of $Q$ for the $k$-th sample drawn for level $l$, determined from a numerical solution on level $\ell$.
Due to the construction of the estimator in~\eqref{hahn:eq_mlmc-scheme},
where we consider the difference in the QoI between consecutive discretization levels for each sample, we recover
\begin{equation}
    \mathbb{E}[\mathbb{E}_{\text{MLMC}}[Q]] = \mathbb{E} [Q_L] \text{,}
\end{equation}
which means that the bias error of the estimation depends on the accuracy of the numerical solution on discretization level $L$, as in the standard MC method.
This approach can also be understood as a variance reduction technique, see e.g.~\cite{Giles_2015aa} for a detailed derivation.
In addition to the total number of samples $N$ as in the MC method, the MLMC method requires us to choose the distribution of those samples across the discretization levels.
If the numerical cost $C_\ell$ associated with one realization of $Q_\ell - Q_{\ell-1}$
and the corresponding variance $V_\ell$  is known for all levels $\ell$, the optimal sample distribution
\begin{equation}
    N_\ell \sim \sqrt{V_\ell / C_\ell}
    \label{hahn:eq_sample_number_distribution}
\end{equation}
follows from an optimization problem~\cite{Giles_2015aa}.
In practice, we may have a good a priori estimation of the numerical cost $C_\ell$, e.g.\ due to the use of hierarchical refinement. The variance of the QoI difference $V_\ell$, however, is not known and has to be estimated at run-time, which necessitates the sequential treatment of discretization levels.
It should be noted that in the limit case of unlimited parallel computing hardware, MLMC sampling cannot reduce the time-to-solution of the estimation  compared to plain MC sampling even if all levels are executed in parallel.
If the sample distribution across levels is not prescribed a priori,
the time-to-solution is even increased due to the sequential treatment of discretization levels. The computational effort, however, is reduced dramatically compared to naive MC sampling.
\subsection{Parallel-in-Time Integration}%
\label{hahn:sec_pint}
To reduce the time-to-solution for numerically expensive models, methods that can make use of parallel computing hardware are desirable.
For models which require small timesteps or the simulation over large timescales, Parallel-in-Time (PinT) methods can become attractive.
The Parareal algorithm, introduced in~\cite{Lions_2001aa} and applied to machines in~\cite{Schops_2018aa},
divides the simulation interval $[T_0, T_M]$ into $M$ sub-intervals $[T_n, T_{n+1}]$ and, in parallel, solves an independent initial value problem (IVP) on each sub-interval
using a time integration method $\mathcal{F}$, e.g.\ the implicit Euler method with time step size $\Delta t_F$, called the fine propagator.
To produce the initial values for the fine propagator, a cheaper and less accurate time integration method $\mathcal{C}$, called the coarse propagator, is used in a sequential pass over the sub-intervals, e.g.\ the implicit Euler method with time step size $\Delta t_C\gg\Delta t_F$.
As the trajectory obtained using the coarse propagator (and therefore the initial values provided to $\mathcal F$) can differ substantially from the one obtained from the fine propagator, this procedure introduces discontinuities at the interval boundaries $T_n$. It is therefore necessary to repeat this process iteratively, until the discontinuities at interval boundaries reach a prescribed tolerance.
Denoting by $\mathbf{U}_n^k$ the initial value used for interval $n$ in iteration $k$, and by $\mathcal{F}(t_m, t_k, \mathbf{U})$ the solution obtained
for $t=t_m$ by the solver $\mathcal{F}$, starting from the initial value $\mathbf{U}$ at time $t_k$,
the Parareal algorithm determines new initial values using the update equation
\begin{equation}
    \begin{aligned}
        \mathbf{U}_n^{(k+1)} &= \mathcal{F}(T_n, T_{n-1}, \mathbf{U}_{n-1}^{(k)}) + \mathcal{C}(T_n, T_{n-1}, \mathbf{U}_{n-1}^{(k+1)}) \\
        &\quad - \mathcal{C}(T_n, T_{n-1}, \mathbf{U}_{n-1}^{(k)}) \text{,}
    \end{aligned}
\end{equation}
in which the fine solution obtained in this iteration, the new coarse solution obtained in a sequential pass in this iteration, and a coarse solution known from the last iteration are combined.
After $k-1$ iterations, the solution on the first $k-1$ intervals is exact up to the accuracy of the fine propagator, and the discontinuities at the first $k-1$ interval boundaries vanish. In iteration $k$, the fine propagator therefore has to be employed for the intervals $k, \,\dots, M$, whereas the coarse propagator
--- being used to generate and update initial values --- is not applied to the last interval, i.e.\ is used on the intervals $k, \,\dots, M-1$.
Using $\tau_{\scriptscriptstyle\mathcal{F}}$ and $\tau_{\scriptscriptstyle\mathcal{C}}$ to denote the numerical cost for a purely sequential solution (over the entire interval) using $\mathcal{F}$ and $\mathcal{C}$, respectively,
we find the contributions of both propagators to the time-to-solution and computational effort in iteration $k$ as
\begin{align}
    \tau_{\text{para}}^{(k)} &= \frac{1}{M} \tau_{\scriptscriptstyle\mathcal{F}} + \frac{M-k}{M} \tau_{\scriptscriptstyle\mathcal{C}} \text{,}\\
    \mathcal{E}_{\text{para}}^{(k)} &= \frac{M-k+1}{M} \tau_{\scriptscriptstyle\mathcal{F}} + \frac{M-k}{M} \tau_{\scriptscriptstyle\mathcal{C}} \text{.}
\end{align}
Ignoring the communication overhead between the different processor cores or machines present in any implementation, we estimate the time-to-solution by
\begin{equation}
  \tau_{\text{para}} = \frac{K}{M} \left[
  \tau_{\scriptscriptstyle\mathcal{F}} +
  \left( M - \frac{K+1}{2} \right) \tau_{\scriptscriptstyle\mathcal{C}}
  \right] \text{,}
  \label{hahn:eq_parareal_time}
\end{equation}
and the computational effort of the Parareal algorithm
\begin{equation}
  \mathcal{E}_{\text{para}} = \frac{K}{M} \left[
  M - \frac{K+1}{2}
  \right]
  \left(
  \tau_{\scriptscriptstyle\mathcal{F}} +\tau_{\scriptscriptstyle\mathcal{C}}
  \right) + \frac{K}{M} \tau_{\scriptscriptstyle\mathcal{F}} \text{,}
  \label{hahn:eq_parareal_effort}
\end{equation}
where $K$ is the number of iterations needed to reach the prescribed tolerance.
The coarse propagator $\mathcal{C}$ has to be chosen carefully,
as it has to be accurate enough to ensure convergence of the algorithm with $K \ll M$. At the same time, $\tau_{\scriptscriptstyle\mathcal{C}}$ should be kept as small as possible to reduce its contributions to~\eqref{hahn:eq_parareal_time} and~\eqref{hahn:eq_parareal_effort}.
As easily observed in~\eqref{hahn:eq_parareal_effort}, using the Parareal algorithm increases the computational effort compared to a sequential solution using $\mathcal{F}$, even if $K\ll M$ and $\tau_{\scriptscriptstyle\mathcal{C}} \ll \tau_{\scriptscriptstyle\mathcal{F}}$ hold.
To achieve a reasonable balance between speedup and computational effort in a UQ setting, the Parareal algorithm should therefore be applied only to relatively few samples.
\subsection{Combined algorithm}%
\label{hahn:sec_proposed}
In the combined MLMC+Parareal method, we aim to use the available computational resources to reduce the time-to-solution of the estimation, while keeping the  computational effort (and thus energy usage) at an acceptable level.
In this setting, we encounter both sample-- and time-parallelism. Since the evaluation of different samples is embarassingly parallel, computational resources should first be used to exploit sample-parallelism.
As described earlier, the variance $V_\ell$ for a given level as required in~\eqref{hahn:eq_sample_number_distribution} is unknown, hindering an up-front determination of the optimal sample distribution across levels. We therefore assume the discretization levels to be treated sequentially.
We expect $N_L$, the number of remaining high-accuracy samples on level $L$ to be small compared to the total number of samples $N = \sum_{\ell=0}^L N_\ell$. In an HPC setting, we also assume $N_L$ as small compared to the number of available processor cores $n_{\text{c}}$, which means that for each sample, additional cores can be leveraged to accelerate the numerical solution.
In the combined method, we therefore restrict the use of Parareal to the finest discretization level $L$,
where we expect the increased computational effort to stay within acceptable bounds due to the small number of samples.

To investigate the performance bounds of the combined method, we first assume a sufficiently parallel computing environment, that is able to perform the numerical solution for all samples on a given level in parallel. The time-to-solution and computational effort needed for an estimation using MLMC without the proposed addition of Parareal then reads
\begin{equation}
  \tau_{\text{ref}, \infty} = \sum_{\ell=0}^{L} C_\ell \text{,} \quad \quad
  \mathcal{E}_{\text{ref}, \infty} = \sum_{\ell=0}^{L} N_\ell C_\ell \text{,}
\end{equation}
whereas the combined method is expected to exhibit
\begin{align}
  \tau_{\text{comb}, \infty} &= \sum_{\ell=0}^{L-1} C_\ell + \tau_{\text{para}} \text{,} \\
  \mathcal{E}_{\text{comb}, \infty} &= \sum_{\ell=0}^{L-1} N_\ell C_\ell + N_L \mathcal{E}_{\text{para}} \text{.}
\end{align}
It should be noted that, even with unlimited parallel computing hardware,
our aim is to bring the time-to-solution $\tau_{\text{para}}$, needed for a single sample on level $L$ using the Parareal algorithm, down to the time needed for a numerical simulation on level $L-1$, i.e.\ $C_{L-1}$.
Any further improvement would be obscured by the
numerical cost associated with samples drawn for level $L-1$. To see a meaningful reduction in the time-to-solution of the full estimation, it would then be necessary to accelerate their solution as well. As the number of samples drawn for level $L-1$ is typically much larger, this is considered infeasible in practice, as it would put greater strain on the computational effort of the method and require a much larger number of available cores.

To achieve this optimal speedup for samples on level $L$, that is to reduce $\tau_{\text{para}}$ down to $C_{L-1}$, it is not sufficient
to choose the coarse propagator to be as costly and as accurate as the (sequential) time integration on the previous level $L-1$, as the initialization of the algorithm would then incur the contribution $\tau_{\scriptscriptstyle\mathcal{C}} = C_{L-1}$ to the time-to-solution already.
Instead, we propose to use the time discretization associated with level $L-2$ for the coarse propagator, i.e.\ $\tau_{\mathcal{C}} = C_{L-2}$.
Assuming an exponential refinement of the time discretization, i.e.\ an exponential decrease of the time step size $\Delta t$ with discretization level, and corresponding cost
\begin{equation}
  C_\ell = C_0 r^\ell
  \label{hahn:eq_exponential_cost}
\end{equation}
for some fixed cost $C_0$ and refinement rate $r$, we find as
\begin{equation}
  s_{\infty}
  = \frac{\sum_{\ell=0}^{L} C_\ell}{\sum_{\ell=0}^{L-1} C_\ell + C_{L-1}} = \frac{r^{L+1}-1}{2r^L - r^{L-1} -1}
\end{equation}
the maximum speedup achievable with the combined method if an infinite number of processor cores is available.
Note that this is the limit case, reached for optimal acceleration of the numerical solution for samples on level $L$.
The number of cores $n_{\text{c, para}}$ needed per sample as well as the number of iterations needed depend on the problem at hand and on the specified tolerance.

The associated increase in computational effort depends most strongly on the number of Parareal iterations $K$ needed to reach convergence, which in turn depends on the problem, the tolerance, and the number of cores $n_{\text{c, para}}$ used per sample.
Demanding $\tau_{\text{para}} \leq C_{L-1}$ for the time-to-solution of the Parareal algorithm in~\eqref{hahn:eq_parareal_time} and using~\eqref{hahn:eq_exponential_cost} for $\ell = L-1$, we find the condition
\begin{equation}
  K
  \leq \kappa
  \leq r \text{,}
\end{equation}
with
\begin{equation*}
  \kappa
  \!=\!\left\lfloor\!\frac{2r^2\!+\!2n_{\text{c, para}}\!-\!1\!-\!\sqrt{(2r^2+2n_{\text{c, para}}-1)^2 -8n_{\text{c, para}} r}}{2}\!\right \rfloor \text{,}
\end{equation*}
on the maximum number of iterations for which the optimal speedup can be achieved, given that $n_{\text{c, para}}$ cores are used for each sample.
This gives us an upper bound $\varepsilon_{\infty, \text{max.}}$ on the increase in computational effort,
\begin{equation}
  \varepsilon_\infty \coloneqq \frac{\mathcal{E}_{\text{comb}, \infty}}{\mathcal{E}_{\text{ref}, \infty}} \leq \varepsilon_{\infty, \text{max}} \text{,}
\end{equation}
where $\varepsilon_{\infty, \text{max}}$ follows from inserting the maximum number of iterations $\kappa$ in the definition of $\varepsilon_{\infty}$.

\begin{figure}[tbh]
    \centering
    \includegraphics{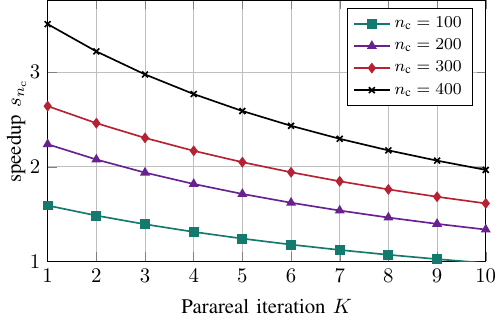}
    \caption{Maximum achievable speedup of the combined method compared to MLMC over the number of Parareal iterations needed on the finest discretization level for various amounts of available processor cores.}%
    \label{hahn:fig_s_nc}
\end{figure}

\begin{figure}[tbh]
  \centering
  \includegraphics{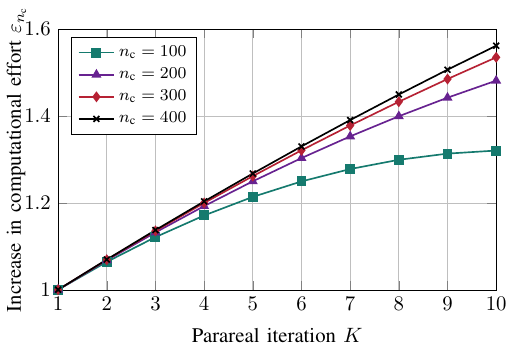}
  \caption{Increase of computational effort of the combined method compared to MLMC over the number of Parareal iterations needed on the finest discretization level for various amounts of available processor cores.}%
  \label{hahn:fig_eps_nc}
\end{figure}

When performing UQ in practice, one does not have an unlimited number of processor cores available, i.e., one cannot assume that all numerical solutions of a given discretization level can be computed in parallel --- rather, it becomes necessary to perform them in batches.
In an optimal implementation, the second numerical solution for each sample --- performed on the previous discretization level --- can be performed in parallel.
With $n_{\text{c}}$ the total number of processor cores available on the system,
\begin{equation}
  b_\ell = \left \lceil \frac{
    N_\ell \left(1  + \frac{C_{\ell-1}}{C_\ell}\right)
  }{n_{\text{c}}} \right \rceil
  \label{hahn:eq_batch_count}
\end{equation}
batches are then needed on level $\ell$, where $\frac{C_{\ell-1}}{C_\ell} = \frac{1}{r}$ if $\ell>0$.
On level $L$, we then have
\begin{equation}
  n_{\text{c, para}} = \left \lfloor \frac{n_{\text{c}}}{
    N_L  \left(1  + \frac{C_{L-1}}{C_L}\right)
  } \right \rfloor
  \label{hahn:eq_num_core_para}
\end{equation}
cores available for each sample, which can be used to accelerate the time integration with the Parareal algorithm.
Combining these assumptions, we find the possible speedup of the estimation as
\begin{equation*}
  s_{n_{\text{c}}}
  = \frac{\sum_{\ell=0}^{L} b_\ell C_\ell}{\sum_{\ell=0}^{L-1} b_\ell C_\ell
  + \frac{K}{n_{\text{c, para}}}
  \left[
  r^L + (n_{\text{c, para}} - \frac{K+1}{2}) r^{L-2}
  \right]} \text{.}
\end{equation*}
Using the previous example of a refinement rate $r=10$, maximum refinement level $L=2$ and $V_\ell \sim 10^{-2\ell}$, Fig.~\ref{hahn:fig_s_nc} shows the possible speedup of the proposed method for different numbers of available processor cores. Figure~\ref{hahn:fig_eps_nc} then shows the associated increase in computational effort.
%%%
%%%
\section{Numerical examples}%
\label{hahn:ch_numerical_example}
To investigate the behavior of the combined method, we apply it to the simulation of two numerical test cases.
First, we consider the Steinmetz equivalent circuit model of an electrical induction machine from~\cite{machine}. The second example is a 2D FE model `im\_3kW' of the induction machine from~\cite{Gyselinck_2000aa} discretized by Gmsh/GetDP~\cite{Geuzaine_2007aa}.
The numerical tests are implemented using the Julia programming language~\cite{julia} and the DifferentialEquations.jl~\cite{diffeq} and MultilevelEstimators.jl~\cite{mlestimators} frameworks.
Parallelization is conducted within Julia, using the Distributed.jl~\cite{distributed} package.

The code and models are publicly available at~\cite{mlmc-parareal-repo}.
All numerical tests were conducted on a single machine with a processor count of $n_{\text{c}} = 180$, but without utilizing shared memory optimizations.
In each case, we compare the time-to-solution and computational effort for an uncertainty quantification between the MLMC method and the proposed combination with Parareal.

\subsection{Steinmetz machine model}
As a first numerical test, we consider the Steinmetz model of the induction machine shown in~\cite[p.\ 108]{machine}, with the machine parameters and their nominal values shown in Tab.~\ref{hahn:tab_nominal-parameters}.
We assume a uniformly distributed variation of $\pm 5 \%$ around the nominal value for each machine parameter.
We perform a transient simulation for $t \in \left[\SI{0}{\second}, \SI{1}{\second}\right]$ using the implicit Euler scheme, with the electrical energy consumed in this timeframe used as the QoI.
The discretization level $\ell$ determines the time step size, with step sizes
$\Delta t \in \{\SI{1e-3}{\second}, \SI{1e-4}{\second}, \SI{1e-5}{\second}\}$
for levels $\ell \in \{0, 1, 2\}$, respectively.
Looking at a single numerical solution using the finest discretization level, we find a speedup of 2.93 when using the Parareal algorithm.
Taking the cost of the coarser samples into account, the overall time-to-solution is reduced by a factor of 1.45
compared to MLMC without Parareal.
The total computational effort is increased by a factor of 1.15.
The speedup observed here is a bit lower than the previous considerations suggest, due to the communication overhead observed in any Parareal implementation. Nevertheless, this numerical test shows that an attractive speedup of the estimation can be achieved with a limited increase in computational effort.
\begin{table}
\caption{Nominal values of machine parameters.}
\begin{tabular}{l|c|l}
    Name & Unit & Value \\ \hline
    Stator resistance & $\SI{}{\ohm}$ & $\SI{1.111140e-01}{}$ \\
    Squirrel cage resistance & $\SI{}{\ohm}$ & $\SI{7.158602e-02}{}$ \\
    Iron loss resistance & $\SI{}{\ohm}$ & $\SI{1.736354e+06}{}$ \\
    Leakage inductance of the stator windings & $\SI{}{\henry}$ & $\SI{1.649983e-03}{}$ \\
    Leakage inductance of the squirrel cage & $\SI{}{\henry}$ & $\SI{1.063014e-03}{}$ \\
    Main inductance & $\SI{}{\henry}$ & $\SI{6.407774e-02}{}$ \\
\end{tabular}%
\label{hahn:tab_nominal-parameters}
\end{table}

\subsection{Finite element model}
As a second test case, we consider a 2D finite element (FE) model of the 4-pole induction machine introduced in~\cite[Ch. 9]{Gyselinck_2000aa}.
Here, the conductivity of each rotor bar is considered to be uncertain, resulting in 32 uncertain parameters. We assume each parameter to follow a uniform distribution of $\pm 5\%$ around the nominal value of $\sigma_{\text{nom.}} = \SI{26.7e6}{\siemens\per\meter}$.
The quantity of interest we consider is the mean joule loss in the simulation time span $\left[0, T\right]$. For a random sample $\boldsymbol{\xi}$, we therefore set
\begin{equation}
  Q(\boldsymbol{\xi}) = \frac{1}{T} \int_{0}^{T} \frac{\mathrm{d}{\mathbf{a}_{\boldsymbol{\xi}}}}{\mathrm{d}{t}} \, \mathbf{M}_{\boldsymbol{\xi}} \, \frac{\mathrm{d}{\mathbf{a}_{\boldsymbol{\xi}}^{\mathrm{T}}}}{\mathrm{d}{t}} \, \mathrm{d}t \text{,}
\end{equation}
where $\mathbf{M}_{\boldsymbol{\xi}}$ depends on $\boldsymbol{\xi}$ directly,
and $\mathbf{a}_{\boldsymbol{\xi}}(t)$ is the solution of~\eqref{hahn:eq_eddy-current-semidiscrete} where $\mathbf{M} = \mathbf{M}_{\boldsymbol{\xi}}$.
We perform a transient simulation over $8$ electrical periods, that is $T = \SI{0.16}{\second}$, and observe the runtimes for a given discretization level shown in Tab.~\ref{hahn:tab_single-runtimes}. Using the Parareal algorithm with $n_{\text{c, para}} = 16$ for a single numerical solution on the finest discretization level $L=2$, we achieve a speedup of $2.69$ compared to the sequential evaluation.
The time-to-solution of the entire estimation is accelerated by a factor of
1.12, at an increase of the computational effort of 1.18.

While the numerical solution for each sample on the finest discretization level benefits from an attractive speedup, the overall reduction of time-to-solution observed here is lower than expected. This is mainly due to the fact that the sample distribution across levels as shown in Tab.~\ref{hahn:tab_single-runtimes} necessitated evaluation in many batches for the samples drawn for $\ell=0$, due to the limited number of available cores of $n_{\text{c}} = 180$. This effect diminishes with increasing core counts, whereas the increase in effort stays constant.
\begin{table}
\caption{Simulation runtime for a single evaluation of the FE model on different discretization levels.}
\begin{tabular}{c|c|c|c|c}
    level $\ell$ & $N_\ell$ & solution mode & step size $\Delta t$ (s) & runtime (s) \\ \hline
    0 & 10303 & Sequential & \SI{2e-3}{} & 10.02 \\
    1 & 85 & Sequential & \SI{2e-4}{} & 47.87 \\
    2 & 10 & Sequential & \SI{2e-5}{} & 397.45 \\
    2 & 10 & Parareal & \SI{2e-5}{} &147.93 \\
\end{tabular}%
\label{hahn:tab_single-runtimes}
\end{table}

\section{Conclusion and Outlook}%
\label{hahn:ch_conclusion}
In this work, we applied the Multi-Level Monte Carlo sampling technique with hierarchical refinement of the temporal resolution to the Uncertainty Quantification of electric machines.
This is a significant advance to previous works --- combining different spatial resolutions --- in cases where long, transient simulations have to be performed.
To accelerate the (usually very costly) Uncertainty Quantification, we proposed a method combining Multi-Level Monte Carlo sampling with parallel-in-time integration, which on its own offers a reduction in the time-to-solution for a single numerical solution, while greatly increasing the computational effort, and thus the energy used to complete the computation.
We investigated the relationship between achievable speedup and increase in energy use for the proposed method, where the speedup depends most strongly on the number of processor cores available. In two numerical test cases, we confirmed the increase in energy usage to stay within an acceptable range of less than $20\%$.
The speedup of $12$ -- $45\%$ observed in the numerical tests is lower than indicated by theoretical considerations, due to previously neglected communication overhead and the limited number of processor cores available for these tests. As the computational effort scales with the number of processor cores much slower than the speedup, we expect a better tradeoff between time-to-solution and energy usage as the number of processor cores increases.
In further research, this assumption shall be tested by applying the proposed method on massively parallel computer systems.
\section*{Acknowledgement}
This work is supported by the Graduate School for CE at TU Darmstadt and the European HPC Joint Undertaking (JU) under grant agreement No. 101118139. The JU receives support from the European Union’s Horizon Europe Programme.

\end{document}